\documentclass[%
 reprint,
 amsmath,amssymb,
 aps,
]{revtex4-2}

\usepackage{graphicx}
\usepackage{dcolumn}
\usepackage{bm}
\usepackage[colorlinks=true,linkcolor=blue,citecolor=blue,urlcolor=blue,]{hyperref}

\begin{document}

\preprint{APS/123-QED}

\title{Kinetic Energy Distribution of Fragments for Thermal Neutron-Induced $^{235}$U and $^{239}$Pu Fission Reactions}
\thanks{This work is supported by National Natural Science Foundations of China (No. 12065003, 12042508, 11465005); Natural Science Foundation of Guangxi (No. 2019GXNSFDA185011) and Key Laboratory of Neutron Physics China Academy of Engineering Physics (No. 2018BA03).}%

\author{Xiaojun Sun$^1$}
 \altaffiliation{sxj0212@gxnu.edu.cn}
\author{Haiyuan Peng$^1$}%
\author{Liying Xie$^1$}
\author{Kai Zhang$^1$}
\author{Yan Liang$^1$}

\author{Yinlu Han$^{1, 2}$}
\author{Nengchuan Shu$^2$}

\author{Jie Yan$^3$}
\author{Jun Xiao$^3$}
\author{Junjie Sun$^3$}

\affiliation{$^1$College of Physics, Guangxi Normal University, Guilin 541004, People's Republic of China}%
\affiliation{$^2$China Institute of Atomic Energy, P. O. Box 275(41), Beijing 102413, People's Republic of China}
\affiliation{$^3$Institute of Nuclear Physics and Chemistry, China Academy of Engineering Physics, Mianyang 621900, People's Republic of China}


\date{\today}

\begin{abstract}
The Dinuclear and Statistical Model (DSM), which focuses on the generation and evolution of vast complementary pairs of the primary fission fragments at the scission moment, is proposed. (1) The fissile nucleus is assumed to elongate along a symmetric coaxis until it breaks into two primary fission fragments. (2) Every complementary pair of the primary fission fragments is approximatively described as two ellipsoids with large deformation. (3) The kinetic energy in the  pair is mainly provided by the Coulomb repulsion, which is explicitly expressed through strict six-dimensional integrals. (4) Only three phenomenological coefficients are obtained to globally describe the quadrupole deformation parameters of arbitrary primary fragments both for $^{235}$U($n_{th}, f$) and $^{239}$Pu($n_{th}, f$) reactions, based on the common characteristics of the measured data, such as mass and charge distributions, kinetic energy distributions. In the framework of DSM, the explicit average total kinetic energy distribution $\overline{TKE}(A)$ and the average kinetic energy distribution $\overline{KE}(A)$ are consistently represented. The theoretical results in this paper agree well with the experimental data. Furthermore, the reliable DSM is expected to generally evaluate the corresponding observables for thermal neutron-induced fission of actinides.

\end{abstract}

\maketitle


\section{INTRODUCTION}\label{sec1}
Nuclear fission has exceptionally challenged the theoretical research since its discovery in the late 1930s \cite{Hahn1939}. The evolution of a nucleus from a compact configuration into two separated fragments is an intricate puzzle \cite{Schmitt2018}, featuring not only the collective movement of large-scale nucleons, but also the various structural effects. The current theoretical descriptions of fission reflect the complexity and richness revealed in experimental studies, emphasizing the multidimensional, dynamic, and microscopic aspects \cite{Vogt2009}. Despite tremendous advances in theory, there is not yet a quantitative theory of fission \cite{Vogt2009}. This is unfortunate because nuclear fission remains important to the society due to its practical applications, in safeguards, accelerator technology, homeland security, medicine, energy production, and waste transmutation at nuclear reactors \cite{Hambsch2014, Neudecker2016, Gooden2016}, and $r$-process in the merging of neutron stars \cite{Goriely2013, Eichler2015}.

The majority of energy released in neutron-induced fission of actinides is in the form of kinetic energy in the fission fragments \cite{Higgins2020}. This kinetic energy is measured by experiments and generally expressed in the relationship with the mass number $A$ of the light and heavy fragments of the average total kinetic energy distribution $\overline{TKE}(A)$ and the average kinetic energy distribution$\overline{KE}(A)$. As an important part of the observables, it has a close relationship with other observables (such as mass distribution, charge distribution, neutron multiplicity, and so on). Moreover, it is closely related to shell effect \cite{Scamps2018}, which is helpful for the research of nuclear structure.

It is universally acknowledged that the transformation of fissile nucleus from a single system to two systems is one of the critical problems. Therefore, a comprehensive description of the deformation of these large primary fragments is indispensable to quantitatively predict the fission products. Some macroscopic models, macroscopic and microscopic models, microscopic models, and time-dependent microscopic theories \cite{Schunck2016, Regnier2016, Zhao2019, Lemaitre2019, Mustonen2018, Ward2017, Goddard2015} have been used to calculate deformation parameters from fissile nucleus to fission fragments. Albeit the well-established physics of these models, the calculated results of fission products vary greatly, with obscure problems concerning the microscopic fission theory. For example, the dissipation coefficient in the fission process is difficult to be calculated by the microscopic method. The evolution relationship between quantum tunneling effect and dissipation effect, and the coupling of different dimensional degrees of freedom in multi-dimensional fission, are necessary to be further considered. It is widely shared that the results of these microscopic models have not yet been adopted by the latest evaluation nuclear data libraries, such as ENDF/B-VIII.0 \cite{Brown2018}, JEFF-3.3.1 \cite{JEFF-3.3.1}, JENDL-4.0u2\cite{JENDL}, CENDL-3.1 \cite{Ge2011}, and so on.

Machine learning method developed in recent years plays a very important role in the evaluation of nuclear data \cite{Niu2018, Wang2019}. It optimizes the theoretical data and the accuracy, but ignores some physical evolution processes. The semi-empirical GEF model \cite{Schmitt2018} summarizes the fundamental laws of physics and the general properties of microscopic systems and mathematical objects. Many fission observables can be more accurately calculated with no need to specifically adjust the measurement and empirical data of a single system. This unique feature, which is of great value in evaluating nuclear data, is difficult to account for the fission process. The pre-scission configuration (PSC) approach represents that the part of the neck will be incorporated into the nascent light and heavy fragments at the scission moments. So the electrostatic interactions between a rotational ellipsoid and a sphere, and between two ellipsoids, are employed to describe the kinetic energy. This approach can provide trust enough $\overline{TKE}(A)$ distributions and extend the fissioning systems for which experimental $\overline{TKE}(A)$ data do not exist \cite{Manea2011}.

In this paper, a new DSM is proposed to concurrently calculate $\overline{TKE}(A)$ and $\overline{KE}(A)$ of fragments for thermal neutron-induced $^{235}$U and $^{239}$Pu fission reactions. In section \ref{sec2}, The derivation process of the Coulomb repulsion is introduced in detail. In section \ref{sec3}, the deformation parameters of the primary fragments at the scission moment are generally described. The calculated results and analyses are shown in section \ref{sec4}. And simple conclusions are given in section \ref{sec5}.

\section{Coulomb repulsion}\label{sec2}
\begin{figure}
\includegraphics[width=8.5cm]{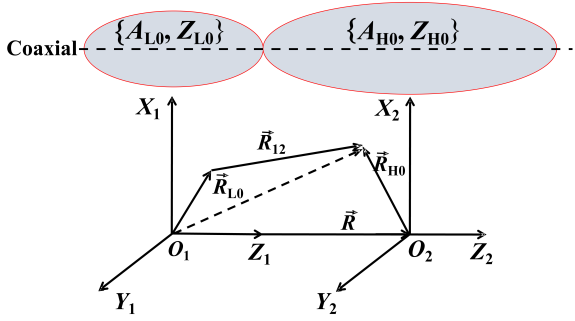}
\caption{\label{fig:FF}The schematic of dinuclear and coordinate system at the scission moment.}
\end{figure}

It is widely shared that the fissile nucleus \{$Z_f, A_f$\} elongates along the symmetric coaxis because of the deformation energy. With its elongation to a certain degree, the fissile nucleus will attain the scission point and split into a multitude of complementary fission fragment pairs \{$Z_{L0}, A_{L0} ; Z_{H0}, A_{H0}$ \} \cite{Madland2006}, which are unstable neutron-rich nuclei and hold large kinetic energy under the Coulomb repulsion. Furthermore, they de-excite through emitting fast neutrons and $\gamma$ photons, and evolve to initial fission products \{$Z_{L0}, A_L; Z_{H0}, A_H$\}, which will further de-excite through emitting slow neutrons and $\beta$ ray to form relatively stable secondary fission products \{$Z_L, A_L; Z_H, A_H$\}.

Complementary primary fragments have complex shapes at the scission moment. Although the expression for the Coulomb interaction of two deformed, arbitrarily oriented, axially symmetric nuclei is obtained  \cite{Denisov2007}, it is difficult to derive the analytical formula. To vividly describ the fissile system at the scission point, the dinuclear concept is used in this paper. If the fissile nucleus is assumed to stretch along the symmetric coaxis until it breaks into two primary fission fragments, every complementary pair of the primary fission fragments is approximatively described by two ellipsoids with large deformation at the scission moment. The schematic of dinuclear and coordinate systems at the scission point is shown in Fig.~\ref{fig:FF}. The kinetic energy in every complementary pair of the primary fragments is mainly provided by the Coulomb repulsion, with the representation of the Coulomb interaction shown as follows \cite{Denisov2007}:
\begin{equation}
\label{Vc}
V_{C}=e^{2} \int \frac{\rho_{L0}\left(\vec{R}_{L0}\right) \rho_{H0}\left(\vec{R}_{H0}\right)}{\left|\vec{R}_{12}\right|} d \vec{R}_{L0} d \vec{R}_{H0},
\end{equation}
the amount of $e^2$ in Eq.~(\ref{Vc}) is a charge constant with a value of 1.44 MeV$\cdot$fm. The distance $\vec{R}_{12}$ between $d\vec{R}_{L0}$ and $d\vec{R}_{H0}$ for the primary fission fragment pairs can be obtained from Fig.~\ref{fig:FF} as
\begin{equation}
\label{R12}
\vec{R}_{12}=\vec{R}+\vec{R}_{H0}-\vec{R}_{L0}.
\end{equation}

The denominator \cite{Denisov2007, Carlson1950} in Eq.~(\ref{Vc}) is
\begin{equation}
\label{1R12}
\begin{aligned}
\frac{1}{{\left| {{{\vec R}_{12}}} \right|}} &= \sum\limits_{{l_{L0}},{l_{H0}} = 0}^\infty  {\frac{{R_{L0}^{{l_{L0}}}R_{H0}^{{l_{H0}}}}}{{{R^{{l_{L0}} + {l_{H0}} + {1}}}}}}\\ &\times \frac{{4\pi {{( - 1)}^{{l_{L0}}}}({l_{L0}} + {l_{H0}})!}}{{\sqrt {(2{l_{L0}} + 1)(2{l_{H0}} + 1)} }}\\
& \times \sum\limits_m {\frac{{{Y_{{l_{L0}}m}}({\theta _{L0}},{\varphi _{L0}})}}{{\sqrt {({l_{L0}} + m)!({l_{L0}} - m)!} }}}\\
& \times \frac{Y_{{l_{H0}}m}({\theta _{H0}},{\varphi _{H0}})} {({l_{H0}} + m)!({l_{H0}} - m)!}.
\end{aligned}
\end{equation}
Where $Y_{lm}(\theta,\varphi)$ is a spherical harmonic function, and $R_{L0}$, $R_{H0}$, $\theta_{L0}$, $\varphi_{L0}$, $\theta_{H0}$, $\varphi_{H0}$ are the spherical coordinates in the laboratory coordinate systems $O_1$ and $O_2$, respectively.

The numerators $\rho_{L0} (\vec{R}_{L0})$ and $\rho_{H0}(\vec{R}_{H0})$ in Eq.~(\ref{Vc}) are the proton density in the light and heavy primary fragments, respectively. Because of the high excited energy of the primary fragments, the proton distribution is assumed to be uniform in their ranges $R_i (\theta_i)$($i=L0$ or $H0$ which denotes the light and heavy primary fragments, as the same in the next text if specifically unmarked). So in this paper, a homogeneous charged drop with a sharp surface is adopted, with its proton density is expressed as
\begin{equation}
\label{rho}
\rho_i (\vec{r})=\left\{\begin{array}{ll}{\rho_i} & {0 \leq r \leq R_i (\theta_i)}\\ 
{0} & {r \geq R_i(\theta_i)}. \end{array}\right.
\end{equation}
$R_i(\theta_i)$ defines the distance from the origin of the coordinate system to the point on the nuclear surface. For an axially deformed system, $R_i(\theta_i)$ is expressed as
\begin{equation}
\label{Rtheta}
R_i(\theta_i)=R_{i}\left[1+\beta_i Y_{lm}(\theta_i)\right],
\end{equation}
where $R_{i}=r_0A_i^{1/3}$, $r_0=1.226$ fm \cite{Wang2013}. $\beta_{i}$ is a quadrupole deformation parameter, which is very important to describe the tensile strength of the primary fragments \cite{Goddard2015}. So the distance between the centers of mass of light and heavy primary fragments can be rewritten as
\begin{equation}
\label{R}
R=R_{L0}\left[1+\beta_{L0} Y_{20}(0)\right]+R_{H0}\left[1+\beta_{H0} Y_{20}(\pi)\right].
\end{equation}

In order to obtain the explicit form of the Coulomb repulsion $V_C$ expressed in Eq. (\ref{Vc}) for different complementary fragment pairs, the charge densities should be firstly calculated through the definition of proton number, i.e. $Z_i=\int \rho_{i}(\vec{r}) d\vec{r}$. Thus, the expression of the light and heavy kernel densities can be represented as
\begin{equation}
\label{rhoLH}
\rho_i=\frac{3 Z_i}{4 \pi R_i^{3}\left(1+\frac{12}{5} B_i^{2}+\frac{16}{35} B_i^{3}\right)},
\end{equation}
where $B_i=\sqrt{5}\beta_i/ \sqrt{16 \pi}$.

Eq.~(\ref{Vc}) requires a six-dimensional integral,  which can be firstly rewritten as:
\begin{equation}
\label{VcQ1}
\begin{aligned}
V_C&=e^2\int \rho_{L0}(\vec{R}_{L0})d\vec{R}_{L0}Q_1.
\end{aligned}
\end{equation}
Where, $Q_1$ can be represented as:
\begin{equation}
\begin{aligned}
Q_1&=\int_{0}^{2\pi}\int_0^{\pi}\sin\theta_{H0}d\theta_{H0}d\varphi_{H0} \\
&\times \int_{0}^{R_{H0}(\theta_{H0})}\rho_{H0}\frac{1}{\left | \vec{R}_{12} \right |}R_{H0}^2dR_{H0}.
\end{aligned}
\end{equation}
The spherical harmonic function $Y_{lm}(\theta,\varphi)$ can be expanded as
\begin{equation}
\label{Ylm}
Y_{lm}(\theta,\varphi)=\sqrt{\frac{(2l+1)(l-m)!}{4\pi(l+m)}}P_{lm}(\cos\theta)e^{im\varphi}.
\end{equation}
If $m\ne0,~ \int_0^{2\pi}e^{im\varphi}d\varphi=0$,  Eq.~(\ref{Ylm}) is meaningless. Conversely, if $m=0, ~\int_0^{2\pi}e^{im\varphi}d\varphi=2\pi$, $Q_1$ can be expressed as:
\begin{equation}
\begin{aligned}
Q_1&=\sqrt{\pi}\rho_{H0}\sum_{l_{L0}=0}^{\infty}\frac{4\pi(-1)^{l_{L0}}R_{L0}^{l_{L0}}}{l_{L0}!R^{l_{L0}+1}}\frac{Y_{l_{L0},0}(\theta_{L0},\varphi_{L0})}{\sqrt{2l_{L0}+1}}\\
&\times \int_{-1}^{1}\sum_{l_{H0}=0}^{\infty}\frac{(l_{L0}+l_{H0})!R_{H0}^{l_{H0}+3}}{l_{H0}!R^{l_{H0}}(l_{H0}+3)}\\
&\times (1-B_{H0}+3B_{H0}x_1^2)^{l_{H0}+3}P_{l_{H0}}(x_1)dx_1,
\end{aligned}
\end{equation}
where $x_1=\cos\theta_{H0}$ and $x_1\in[-1,~1]$.

Obviously, the integral and summation can be exchanged. Thus, $Q_1$ can be rewritten as
\begin{equation}
\label{Q1}
\begin{aligned}
Q_1&=\sqrt{\pi}\rho_{H0}\sum_{l_{L0}=0}^{\infty}\frac{4\pi(-1)^{l_{L0}}R_{L0}^{l_{L0}}Y_{l_{L0},0}(\theta_{L0},\varphi_{L0})}{l_{L0}!R^{l_{L0}+1}\sqrt{2l_{L0}+1}}\\
&\times [l_{L0}!p_0+(l_{L0}+2)!p_2+(L_{L0}+4)!p_4\\
&+(l_{L0}+6)!p_6+\dots \dots ].
\end{aligned}
\end{equation}
While $l_{H0}=0,~2,~4,~6$, and $p_0,~p_2,~p_4,~p_6$ are expressed as follows
\begin{equation}
\label{p}
\begin{aligned}
p_0&=\frac{R_{H0}^3}{3}(2+\frac{24}{5}B_{H0}^2+\frac{32}{35}B_{H0}^3),\\
p_2&=\frac{R_{H0}^5}{10R^2}(4B_{H0}+\frac{32}{7}B_{H0}^2+\frac{96}{7}B_{H0}^3\\
&+\frac{640}{77}B_{H0}^4+\frac{3392}{1001}B_{H0}^5),\\
p_4&=\frac{R_{H0}^7}{168R^4}(\frac{48}{5}B_{H0}^2+\frac{192}{11}B_{H0}^3+\frac{6528}{143}B_{H0}^4\\
&+\frac{6144}{143}B_{H0}^5+\frac{5376}{187}B_{H0}^6+\frac{334848}{46189}B_{H0}^6),\\
p_6&=\frac{R_{H0}^9}{6480R^6}(\frac{3456}{143}B_{H0}^3+\frac{41472}{715}B_{H0}^4\\
&+\frac{373248}{2431}B_{H0}^5+\frac{8736768}{46189}B_{H0}^6\\
&+\frac{7796736}{46189}B_{H0}^7+\frac{85598208}{1062347}B_{H0}^8\\
&+\frac{467361792}{26558675}B_{H0}^9).
\end{aligned}
\end{equation}

By substituting Eqs. (\ref{Q1}) and (\ref{p}) into Eq. (\ref{VcQ1}), $V_C$ can be rewritten as:
\begin{equation}
\begin{aligned}
V_C=e^2\rho_{L0}\rho_{H0}\sqrt{\pi}Q_2.
\end{aligned}
\end{equation}
Where $Q_2$ can be expressed as:
\begin{equation}
\begin{aligned}
Q_2&=\sqrt{\pi}\int_{-1}^{1}\sum_{l_{L0}=0}^{\infty}\frac{4\pi(-1)^{l_{L0}}R_{L0}^{l_{L0}+3}}{(l_{L0}+3)l_{L0}!R^{l_{L0}+1}}\\
&\times [l_{L0}!p_0+(l_{L0}+2)!p_2+(l_{L0}+4)!p_4\\
&+(l_{L0}+6)!p_6+\dots](1-B_{L0}+3B_{L0}x_2^2)^{l_{L0}+3}\\
&\times P_{l_{L0}}(x_2)dx_2.
\end{aligned}
\end{equation}
Where $x_2=\cos\theta_{L0}$ and $x_2\in[-1,1]$.

As same as $Q_1$, $Q_2$ can be rewritten as:
\begin{equation}
Q_2=\sqrt{\pi}(s_0+s_2+s_4+s_6+\dots).
\end{equation}
While $l_{L0}=0, 2, 4, 6$, $s_0, s_2, s_4, s_6$ are expressed as follows:
\begin{equation}
\begin{aligned}
s_0&=\frac{4\pi R_{L0}^3}{3R}(p_0+2!p_2+4!p_4+6!p_6+\dots)\\
&\times (2+\frac{24}{5}B_{L0}^2+\frac{32}{35}B_{L0}^3),\\
s_2&=\frac{4\pi R_{L0}^5}{3R}\frac{3R_{L0}}{10R^2}(2!p_0+4!p_2+6!p_4+8!p_6+\dots)\\
&\times (4B_{L0}+\frac{32}{7}B_{L0}^2+\frac{96}{7}B_{L0}^3+\frac{640}{77}B_{L0}^4\\
&+\frac{3392}{1001}B_{L0}^5),\\
s_4&=\frac{4\pi R_{L0}^3}{3R}\frac{3R_{L0}^4}{168R^4}(4!p_0+6!p_2+8!p_4\\
&+10!p_6+\dots)(\frac{48}{5}B_{L0}^2+\frac{192}{11}B_{L0}^3+\frac{6528}{143}B_{L0}^4\\
&+\frac{6144}{143}B_{L0}^5+\frac{5376}{187}B_{L0}^6+\frac{334848}{46189}B_{L0}^7),\\
s_6&=\frac{4\pi R_{L0}^3}{3R}\frac{3R_{L0}^6}{6480R^6}(6!p_0+8!p_2+10!p_4\\
&+12!p_6+\dots)(\frac{3456}{143}B_{L0}^3+\frac{41472}{715}B_{L0}^4\\
&+\frac{373248}{2431}B_{L0}^5+\frac{8736768}{46189}B_{L0}^6+\frac{7796736}{46189}B_{L0}^7\\
&+\frac{85598208}{1062347}B_{L0}^8+\frac{467361792}{26558675}B_{L0}^9).\\
\end{aligned}
\end{equation}

After sorting out and omitting the higher order terms, the explicit Coulomb repulsion can be obtained:
\begin{equation}
\label{Vc1}
\begin{aligned}
V_C&=\frac{e^2Z_{L0}Z_{H0}}{Rf_0(B_{L0})f_0(B_{H0})}\{[f_0(B_{H0})\\
&+2!f_2(B_{H0},R_{H0})+4!f_4(B_{H0},R_{H0})]f_0(B_{L0})\\
&+[2!f_0(B_{H0})+4!f_2(B_{H0},R_{H0})\\
&+6!f_4(B_{H0},R_{H0})]f_2(B_{L0},R_{L0})\\
&+[4!f_0(B_{H0})+6!f_2(B_{H0},R_{H0})\\
&+8!f_4(B_{H0},R_{H0})]f_4(B_{L0},R_{L0})\}.
\end{aligned}
\end{equation}
Where the compact forms of $f$ function are listed as follows:
\begin{equation}
\label{f}
\begin{aligned}
f_0 (x)&=1+\frac{12}{5}x^2+\frac{16}{35}x^3,\\
f_2(x,y)&=\frac{3y^2}{10R^2}(2x+\frac{16}{7}x^2+\frac{48}{7}x^3\\
&+\frac{320}{77}x^4+\frac{1696}{1001}x^5),\\
f_4(x,y)&=\frac{3y^4}{168R^4}(\frac{24}{5}x^2+\frac{96}{11}x^3\frac{3264}{143}x^4\\
&+\frac{3072}{143}x^5+\frac{2688}{187}x^6+\frac{167424}{46189}x^7).
\end{aligned}
\end{equation}

Thus, the total kinetic energy of every complementary primary fragment pair can be rewritten as:
\begin{equation}
TKE=V_C(A_{L0}, Z_{L0}, \beta_{L0}; A_{H0}, Z_{H0}, \beta_{H0}).
\end{equation}
And the average total kinetic energy of the complementary primary fragment pairs is expressed as:
\begin{equation}
\label{atke}
\begin{aligned}
\overline{TKE}(A)=&\frac{1}{\sum_{k}}\sum_{k}TKE(A_{L0}(j), Z_{L0}(j, k),\\ &\beta_{L0}(j, k);
 A_{H0}(j), Z_{H0}(j, k),\\ & \beta_{H0}(j, k)),
\end{aligned}
\end{equation}
where $j$ denotes the number of the primary fragment pairs, and $k$ denotes the isobar numbers of the $j$-th primary fragment pair.

From the Eqs. (\ref{Vc1}) - (\ref{atke}), the deformation parameters $\beta_{L0, H0}$ can be seen indispensable. However, it is impossible that these parameters can be experimentally and theoretically derived from in an accurate way. So in this paper, these
quantities are obtained by the following methods in the next sections.

\section{deformation parameters}\label{sec3}
\subsection{Most Probable Primary Fragment Pair}

\begin{figure}
\includegraphics[width=8.5cm]{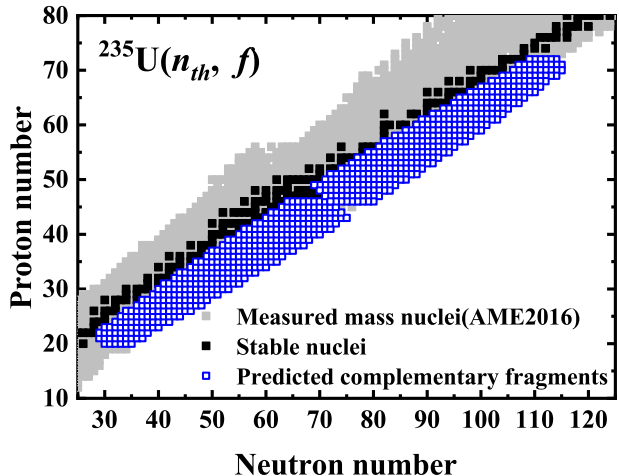}
\caption{\label{fig:PFD}(Color online) Possible primary fragment distribution for $^{235}$U($n_{th},~f$) fission reaction.}
\end{figure}

\begin{figure}
\includegraphics[width=8.5cm]{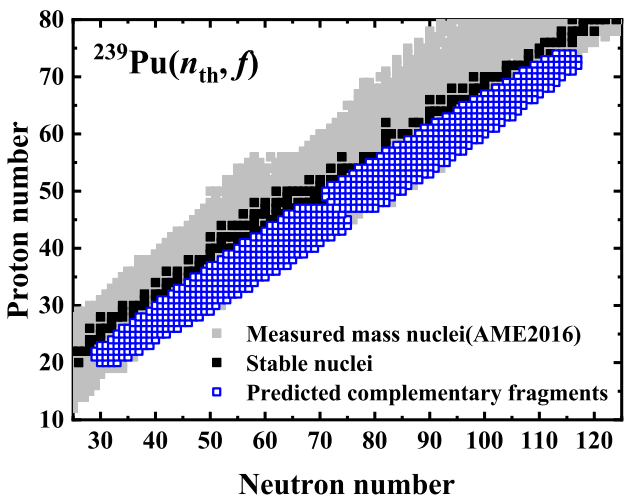}
\caption{\label{fig:PFDPu}(Color online) The same as Fig. \ref{fig:PFD} but for $^{239}$Pu($n_{th}, f$) reaction.}
\end{figure}

It is widely shared that the primary fragment pairs at the scission moment must follow the laws: 
\begin{equation}
\label{laws}
\begin{aligned}
Z_f &=Z_{L0}+Z_{H0},\\
A_f &=A_{L0}+A_{H0}.
\end{aligned}
\end{equation}
In addition, the neutron separation energy of every primary fragment must hold the positive value, i.e., $S_n(Z_i, A_i)>0$. Furthermore, the neutron-proton ratio of every possible primary fragment is larger than that of the fissile nucleus and the corresponding beta-decay stable nuclei. Thus, all of the possible primary fragment pairs can be predicted, as shown in Figs.~\ref{fig:PFD} and \ref{fig:PFDPu} for 
$^{235}$U($n_{th}, f$) and $^{239}$Pu($n_{th}, f$) reactions, respectively. In thses two figures, the gray points denote the measured mass nulcei compiled in AME2016 \cite{Wang2017}, while the black points denote the stable nuclei localed in the vicinity of the beta stable line and the blue hollow points indicate the possible primary fragments predicted in this paper. Obviously, there are hundreds of possible primary fragment pairs \{$A_{L0}(j), Z_{L0}(j, k); A_{H0}(j), Z_{H0}(j, k)$\}, where $j$ denotes the sequence number of the primary fragment pairs, and $k$ labels the sequence number of the isobar pairs for the $j$-th primary fragment pair. From Figs. \ref{fig:PFD} and \ref{fig:PFDPu}, slight discrepancies can be seen at the positions of the symmetrical fission points \{$A_f/2, Z_f/2$\} for $^{235}$U($n_{th}, f$) and $^{239}$Pu($n_{th}, f$) reactions, respectively. And the total primary fragment amount of $^{239}$Pu($n_{th}, f$) reaction is a bit fewer than that of $^{235}$U($n_{th}, f$)reaction, which is slightly unexpected because the fissile nucleus with larger mass are generally believed to produce more primary fragments.

Theoretically, each primary fragment pair has different contribution to the kinetic energy. However,  $\overline{TKE}(A)$ and $\overline{KE}(A)$ for fission reactions are only the functions of the mass number. How to describe the partial contributions of each isobar pairs to the same fragment mass number $A$ is one of the critical problems. To solve it, the common characteristics of the deformation for the vast primary fragments must be revealed. In this paper, it is assumed that only a pair of isobar, named after Most Probable Primary Fragment Pair, dominates the contributions to the same mass number $A$ of the primary fragment, and the excited energies of the primary fragments are not high because of the large deformation energies at the scission moment. Thus, the Most Probable Primary Fragment Pairs can be selected on the basis of the characteristics of the ground state (such as half-life) of the primary fragment pairs. 

\begin{table}
\caption{\label{tab:t} Partial possible primary fragment pairs and their half-lives in the ground states for $^{235}$U$(n_{th},~f)$ fission reaction. The experimental data are taken from IAEA - Nuclear Data Section \cite{IAEA}. Herein, units such as s, ms, min, h, d and y denote second, millisecond, minute, hour, day and year, respectively.}
\begin{ruledtabular}
\begin{tabular}{cccccccc}
$j$&$k$&$A_{L0}(j)$&$Z_{L0}(j, k)$&$A_{H0}(j)$&$Z_{H0}(j, k)$&$\tau_{L0}$&$\tau_{H0}$ \\\hline
1&1&118&	46& 118&	46	&1.9s	  &  1.9s     \\
1&2&118&	45& 118&	47	&266ms	&  3.76s    \\
1&3&118&	44& 118&	48	&99ms	  &  2.69min  \\
1&4&118&	43& 118&	49	&30ms	  &  5s       \\
2&1&117&	46& 119&	46	&4.3s	  &  0.92s    \\
2&2&117&	45& 119&	47	&0.44s	&  2.1s     \\
2&3&117&	44& 119&	48	&151ms	&  2.69min  \\
2&4&117&	43& 119&	49	&44.5ms	&  2.4min   \\
3&1&116&	46& 120&	46	&11.8s	&  0.492s   \\
3&2&116&	45& 120&	47	&0.68s	&  1.23s    \\
3&3&116&	44& 120&	48	&204ms	&  50.8s    \\
3&4&116&	43& 120&	49	&57ms	  &  3.08s    \\
4&1&115&	46& 121&	46	&25s	  &  0.285s   \\
4&2&115&	45& 121&	47	&0.99s	&  0.78s    \\
4&3&115&	44& 121&	48	&0.318s	&  13.5s    \\
4&4&115&	43& 121&	49	&78ms	  &  23.1s    \\
4&5&115&	42& 121&	50	&45.5ms	&  27.03h   \\
5&1&114&	46& 122&	46	&2.42min&	0.108s    \\
5&2&114&	45& 122&	47	&1.85s	&  0.529s   \\
5&3&114&	44& 122&	48	&0.54s  &  5.24s    \\
5&4&114&	43& 122&	49	&0.1s	  &  6.17s    \\
5&5&114&	42& 122&	50	&58ms	  &  Stable   \\
6&1&113&	46& 123&	46	&93s	  &  0.108s   \\
6&2&113&	45& 123&	47	&2.8s	  &  0.298s   \\
6&3&113&	44& 123&	48	&0.8s	  &  2.1s     \\
6&4&113&	43& 123&	49	&0.152s	&  6.17s    \\
6&5&113&	42& 123&	50	&80ms	  &  129.2d   \\
6&6&113&	41& 123&	51	&32ms	  &  Stable   \\
7&1&112&	46& 124&	46	&21.04h	&  38ms     \\
7&2&112&	45& 124&	47	&3.6s	  &  0.191s   \\
7&3&112&	44& 124&	48	&1.75s	&  1.25s    \\
7&4&112&	43& 124&	49	&0.271s	&  3.12s    \\
7&5&112&	42& 124&	50	&0.12s	&  Stable   \\
7&6&112&	41& 124&	51	&33ms	  &  60.2d    \\
8&1&111&	46& 125&	46	&23.4min&	57ms      \\
8&2&111&	45& 125&	47	&11s	  &  0.159s   \\
8&3&111&	44& 125&	48	&2.12s	&  0.68s    \\
8&4&111&	43& 125&	49	&0.29s	&  2.36s    \\
8&5&111&	42& 125&	50	&0.186s	&  9.64d    \\
8&6&111&	41& 125&	51	&54ms	  &  2.76y    \\
9&1&110&	46&	126&	46&	Stable	&48.6ms \\
9&2&110&	45&	126&	47&	3.35s	  &0.052s \\
9&3&110&	44&	126&	48&	12.04s	&0.515s \\
9&4&110&	43&	126&	49&	0.9s	  &1.53s  \\
9&5&110&	42&	126&	50&	0.296s	&2.3$\times 10^5$y\\
9&6&110&	41&	126&	51&	0.082s	&12.35d \\
9&7&110&	40&	126&	52&	37.5ms	&Stable \\
10&1&109&	45&	127&	47&	80.8s	  &0.109s \\
10&2&109&	44&	127&	48&	34.4s	  &0.37s  \\
10&3&109&	43&	127&	49&	0.91s	  &1.09s  \\
10&4&109&	42&	127&	50&	0.61s	  & 2.1h  \\
10&5&109&	41&	127&	51&	0.108s	&3.85d  \\
10&6&109&	40&	127&	52&	0.056s	&9.35h  \\
\end{tabular}
\end{ruledtabular}
\end{table}

\begin{table}
\caption{\label{tab:t_Pu} Partial possible primary fragment pairs and their half-lives in the ground states for $^{239}$Pu$(n_{th},~f)$ reaction. }
\begin{ruledtabular}
\begin{tabular}{cccccccc}
$j$&$k$&$A_{L0}(j)$&$Z_{L0}(j, k)$&$A_{H0}(j)$&$Z_{H0}(j, k)$&$\tau_{L0}$&$\tau_{H0}$ \\\hline
1&1&120	  &47	&120	&47	&1.23s	&1.23s  \\
1&2&120	  &46	&120	&48	&492ms	&50.8s  \\
1&3&120	  &45	&120	&49	&132ms	&3.08s  \\ 
2&1&119	  &47	&121	&47	&2.1s	  &0.78s \\
2&2&119	  &46	&121	&48	&0.92s	&13.5s \\
2&3&119	  &45	&121	&49	&171ms	&23.1s  \\ 
2&4&119	  &44	&121	&50	&69.5ms	&27.03h\\
3&1&118	  &47	&122	&47	&3.76s	&0.529s \\ 
3&2&118	  &46	&122	&48	&1.9s	  &5.24s \\ 
3&3&118	  &45	&122	&49	&266ms	&1.5s  \\
3&4&118	  &44	&122	&50	&99ms	  &Stable \\ 
4&1&117	  &47	&123	&47	&72.8s	&0.298s   \\  
4&2&117	  &46	&123	&48	&4.3s	  &2.1s  \\
4&3&117	  &45	&123	&49	&0.44s	&6.17s    \\
4&4&117	  &44	&123	&50	&151ms	&129.2d  \\
4&4&117	  &43	&123	&51	&44.5ms	&Stable  \\
5&1&116	  &47	&124	&47	&230s	  &191ms \\
5&2&116	  &46	&124	&48	&11.8s	&1.25s   \\ 
5&3&116	  &45	&124	&49	&0.68s	&3.12s\\
5&4&116	  &44	&124	&50	&204ms	&Stable  \\
5&5&116	  &43	&124	&51	&57ms	  &60.2d  \\ 
6&1&115	  &47	&125	&47	&20.0min	&159ms \\  
6&2&115	  &46	&125	&48	&25s	  &0.68s  \\ 
6&3&115	  &45	&125	&49	&0.99s	&2.36s  \\ 
5&4&115	  &44	&125	&50	&318ms	&9.64d   \\
6&5&115	  &43	&125	&51	&78ms	  &2.75856y\\
7&1&114	  &47	&126	&47	&4.6s	  &52ms  \\
7&2&114	  &46	&126	&48	&2.42min	&0.515s  \\ 
7&3&114	  &45	&126	&49	&1.85s	&1.53s  \\   
7&4&114	  &44	&126	&50	&0.54s	&2.3$\times 10^5$y \\
7&5&114	  &43	&126	&51	&100ms	&12.35d  \\ 
7&6&114	  &42	&126	&52	&58ms	  &Stable \\ 
8&1&113	  &47	&127	&47	&5.37h	&109ms  \\ 
8&2&113	  &46	&127	&48	&93s	  &0.37s  \\
8&3&113	  &45	&127	&49	&2.8s	  &1.09s\\
8&4&113	  &44	&127	&50	&0.8s	  &2.1h   \\ 
8&5&113	  &43	&127	&51	&152ms	&3.85d   \\
8&6&113	  &42	&127	&52	&80ms	  &9.35h  \\ 
9&1&112	  &46	&128	&48	&21.04h	&0.28s   \\  
9&2&112	  &45	&128	&49	&3.6s	  &0.84s \\  
9&3&112	  &44	&128	&50	&1.75s	&59.07min \\  
9&4&112	  &43	&128	&51	&271ms	&9.05h   \\
9&5&112	  &42	&128	&52	&120ms	&7.7$\times 10^24$y\\
10&1&111	&46	&129	&48	&23.4min	&154ms \\
10&2&111	  &45	&129	&49	&11s	  &611ms  \\
10&3&111	  &44	&129	&50	&2.12s	&2.23min \\     
10&4&111	  &43	&129	&51	&290ms	&4.366h  \\
10&5&111	  &42	&129	&52	&186ms	&69.6min  \\
10&6&111	&41	&129	&53	&54ms	  &1.57$\times 10^7$y\\
\end{tabular}
\end{ruledtabular}
\end{table}

Tables \ref{tab:t} and \ref{tab:t_Pu} partially list the possible primary fragment pairs and their half-lives in the ground states for $^{235}$U($n_{th},~f$) and $^{239}$Pu($n_{th},~f$) reactions, respectively. The experimental data are taken from IAEA - Nuclear Data Section \cite{IAEA}, with the first column denoting the $j$-th primary fragment pair, the second column labeling the $k$-th isobar pair of the $j$-th primary fragment pair, and the last two columns indicating the half-lives of the light and heavy primary fragments, respectively. Herein, the units such as s, ms, min, h, d and y, denote second, millisecond, minute, hour, day and year, respectively. From tables \ref{tab:t} and \ref{tab:t_Pu}, the half-lives of the light and heavy primary fragments can be seen to exhibit large discrepancies. In this paper, the isobar pairs with the smallest values $\tau_{L0}+\tau_{H0}$ are assumed to be the candidates of the Most Probable Primary Fragment Pairs, all of which are listed in Tables \ref{tab:beta} and \ref{tab:beta_Pu} for $^{235}$U($n_{th},~f$) and $^{239}$Pu($n_{th},~f$) reactions, respectively. It is worth mentioning that there are several isobar pairs, which values $\tau_{L0}+\tau_{H0}$ are roughly equal for some primary fragment pairs. Therefore, the isobar pairs with the smallest values of $|\tau_{L0} - \tau_{H0}|$ are empirically selected as the candidates of the Most Probable Primary Fragment Pairs. 

\begin{table*}
\caption{\label{tab:beta}The properties of the Most Probable Primary Fragment Pairs for $^{235}$U($n_{th},~f$) reaction. $\beta_i, \alpha_i$ and $I_i$ denote the quadrupole deformation parameter, phenomenological parameter and the isospin asymmetry degree, respectively.}
\begin{ruledtabular}
\begin{tabular}{cccccccccc}
$A_{L0}$&$Z_{L0}$&$\beta_{L0}$&$\alpha_{L0}$&$I_{L0}$&$A_{H0}$&$Z_{H0}$&$\beta_{H0}$
&$\alpha_{H0}$&$I_{H0}$ \\\hline
118&	46&	 1.3445& 6.1020&0.2203 &118&	46&  1.3445& 6.1020	&0.2203 \\   
117&	45&	 1.3796& 5.9782&0.2308 &119&	47&  1.2559& 5.9782	&0.2101 \\   
116&	45&	 1.3147& 5.8658&0.2241 &120&	47&  1.2709& 5.8658	&0.2167 \\   
115&	45&	 1.2599& 5.7955&0.2174 &121&	47&  1.2932& 5.7955	&0.2231 \\   
114&	45&	 1.1506& 5.4652&0.2105 &122&	47&  1.2543& 5.4652	&0.2295 \\   
113&	45&	 1.0701& 5.2574&0.2035 &123&	47&  1.2395& 5.2574	&0.2358 \\   
112&	44&	 1.0736& 5.0099&0.2143 &124&	48&  1.1313& 5.0099	&0.2258 \\   
111&	43&	 0.9958& 4.8058&0.2072 &125&	48&  1.1149& 4.8058	&0.2320 \\   
110&	43&	 0.9881& 4.5287&0.2182 &126&	49&  1.0064& 4.5287	&0.2222 \\   
109&	43&	 0.9204& 4.3618&0.2110 &127&	49&  0.9960& 4.3618	&0.2283 \\   
108&	43&	 0.8575& 4.2094&0.2037 &128&	49&  0.9866& 4.2094	&0.2344 \\   
107&	43&	 0.8157& 4.1561&0.1963 &129&	49&  0.9987& 4.1561	&0.2403 \\   
106&	43&	 0.7882& 4.1777&0.1887 &130&	49&  1.0284& 4.1777	&0.2462 \\   
105&	42&	 0.8241& 4.1205&0.2000 &131&	50&  0.9751& 4.1205	&0.2366 \\   
104&	42&	 0.7968& 4.1435&0.1923 &132&	50&  1.0045& 4.1435	&0.2424 \\   
103&	42&	 0.7778& 4.2166&0.1845 &133&	50&  1.0462& 4.2166	&0.2481 \\   
102&	41&	 0.8281& 4.2231&0.1961 &134&	51&  1.0085& 4.2231	&0.2388 \\   
101&	41&	 0.8226& 4.3727&0.1881 &135&	51&  1.0689& 4.3727	&0.2444 \\   
100&	41&	 0.8214& 4.5633&0.1800 &136&	51&  1.1408& 4.5633	&0.2500 \\   
 99&	40&	 0.8834& 4.6031&0.1919 &137&	52&  1.1088& 4.6031	&0.2409 \\   
 98&	41&	 0.7938& 4.8619&0.1633 &138&	51&  1.2683& 4.8619	&0.2609 \\   
 97&	39&	 0.9181& 4.6870&0.1959 &139&	53&  1.1127& 4.6870	&0.2374 \\   
 96&	39&	 0.9086& 4.8460&0.1875 &140&	53&  1.1769& 4.8460	&0.2429 \\   
 95&	38&	 0.9673& 4.8368&0.2000 &141&	54&  1.1320& 4.8368	&0.2340 \\   
 94&	37&	 1.0198& 4.7930&0.2128 &142&	55&  1.0801& 4.7930	&0.2254 \\   
 93&	37&	 0.9998& 4.8938&0.2043 &143&	55&  1.1293& 4.8938	&0.2308 \\   
 92&	37&	 0.9780& 4.9986&0.1957 &144&	55&  1.1802& 4.9986	&0.2361 \\   
 91&	36&	 1.0249& 4.9087&0.2088 &145&	56&  1.1172& 4.9087	&0.2276 \\   
 90&	35&	 1.0752& 4.8382&0.2222 &146&	57&  1.0604& 4.8382	&0.2192 \\   
 89&	35&	 1.0652& 4.9898&0.2135 &147&	57&  1.1202& 4.9898	&0.2245 \\   
 88&	35&	 1.0532& 5.1492&0.2045 &148&	57&  1.1829& 5.1492	&0.2297 \\   
 87&	34&	 1.0945& 5.0114&0.2184 &149&	58&  1.1099& 5.0114	&0.2215 \\   
 86&	33&	 1.1339& 4.8756&0.2326 &150&	59&  1.0401& 4.8756	&0.2133 \\   
 85&	33&	 1.1210& 5.0151&0.2235 &151&	59&  1.0960& 5.0151	&0.2185 \\   
 84&	33&	 1.1080& 5.1708&0.2143 &152&	59&  1.1566& 5.1708	&0.2237 \\   
 83&	33&	 1.0947& 5.3449&0.2048 &153&	59&  1.2227& 5.3449	&0.2288 \\   
 82&	33&	 1.0665& 5.4659&0.1951 &154&	59&  1.2778& 5.4659	&0.2338 \\   
 81&	32&	 1.1248& 5.3592&0.2099 &155&	60&  1.2101& 5.3592	&0.2258 \\   
 80&	33&	 1.0229& 5.8451&0.1750 &156&	59&  1.4238& 5.8451	&0.2436 \\   
 79&	31&	 1.1504& 5.3461&0.2152 &157&	61&  1.1918& 5.3461	&0.2229 \\   
 78&	31&	 1.1440& 5.5771&0.2051 &158&	61&  1.2707& 5.5771	&0.2278 \\   
 77&	30&	 1.2168& 5.5116&0.2208 &159&	62&  1.2132& 5.5116	&0.2201 \\   
 76&	30&	 1.1706& 5.5601&0.2105 &160&	62&  1.2510& 5.5601	&0.2250 \\   
\end{tabular}
\end{ruledtabular}
\end{table*}

\begin{table*}
\caption{\label{tab:beta_Pu}The properties of the Most Probable Primary Fragment Pairs for $^{239}$Pu($n_{th},~f$) reaction.}
\begin{ruledtabular}
\begin{tabular}{cccccccccc}
$A_{L0}$&$Z_{L0}$&$\beta_{L0}$&$\alpha_{L0}$&$I_{L0}$&$A_{H0}$&$Z_{H0}$&$\beta_{H0}$
&$\alpha_{H0}$&$I_{H0}$ \\\hline
120	&47	&1.3458	&6.2112	&0.2167	&120	&47	&1.3458	&6.2112	&0.2167 \\
119	&47	&1.2634	&6.0140	&0.2101	&121	&47	&1.3420	&6.0140	&0.2231 \\
118	&45	&1.2931	&5.4494	&0.2373	&122	&49	&1.0720	&5.4494	&0.1967 \\
117	&46	&1.1439	&5.3533	&0.2137	&123	&48	&1.1751	&5.3533	&0.2195 \\
116	&45	&1.1684	&5.2129	&0.2241	&124	&49	&1.0930	&5.2129	&0.2097 \\
115	&45	&1.0707	&4.9252	&0.2174	&125	&49	&1.0638	&4.9252	&0.2160 \\
114	&45	&1.0014	&4.7567	&0.2105	&126	&49	&1.0570	&4.7567	&0.2222 \\
113	&45	&0.9579	&4.7061	&0.2035	&127	&49	&1.0746	&4.7061	&0.2283 \\
112	&45	&0.9147	&4.6567	&0.1964	&128	&49	&1.0914	&4.6567	&0.2344 \\
111	&45	&0.8628	&4.5606	&0.1892	&129	&49	&1.0960	&4.5606	&0.2403 \\
110	&45	&0.8236	&4.5300	&0.1818	&130	&49	&1.1151	&4.5300	&0.2462 \\
109	&44	&0.8661	&4.4956	&0.1927	&131	&50	&1.0639	&4.4956	&0.2366 \\
108	&45	&0.7658	&4.5945	&0.1667	&132	&49	&1.1834	&4.5945	&0.2576 \\
107	&43	&0.8834	&4.5011	&0.1963	&133	&51	&1.0491	&4.5011	&0.2331 \\
106	&43	&0.8615	&4.5661	&0.1887	&134	&51	&1.0904	&4.5661	&0.2388 \\
105	&42	&0.9075	&4.5377	&0.2000	&135	&52	&1.0420	&4.5377	&0.2296 \\
104	&42	&0.8851	&4.6024	&0.1923	&136	&52	&1.0829	&4.6024	&0.2353 \\
103	&41	&0.9401	&4.6111	&0.2039	&137	&53	&1.0434	&4.6111	&0.2263 \\
102	&41	&0.9134	&4.6585	&0.1961	&138	&53	&1.0802	&4.6585	&0.2319 \\
101	&41	&0.9094	&4.8341	&0.1881	&139	&53	&1.1477	&4.8341	&0.2374 \\
100	&41	&0.8943	&4.9685	&0.1800	&140	&53	&1.2066	&4.9685	&0.2429 \\
99	&40	&0.9468	&4.9332	&0.1919	&141	&54	&1.1546	&4.9332	&0.2340 \\
98	&39	&0.9980	&4.8902	&0.2041	&142	&55	&1.1020	&4.8902	&0.2254 \\
97	&39	&0.9818	&5.0122	&0.1959	&143	&55	&1.1567	&5.0122	&0.2308 \\
96	&39	&0.9639	&5.1406	&0.1875	&144	&55	&1.2138	&5.1406	&0.2361 \\
95	&38	&1.0098	&5.0489	&0.2000	&145	&56	&1.1491	&5.0489	&0.2276 \\
94	&37	&1.0646	&5.0038	&0.2128	&146	&57	&1.0967	&5.0038	&0.2192 \\
93	&37	&1.0380	&5.0805	&0.2043	&147	&57	&1.1405	&5.0805	&0.2245 \\
92	&37	&1.0326	&5.2778	&0.1957	&148	&57	&1.2125	&5.2778	&0.2297 \\
91	&36	&1.0708	&5.1287	&0.2088	&149	&58	&1.1359	&5.1287	&0.2215 \\
90	&35	&1.1234	&5.0553	&0.2222	&150	&59	&1.0785	&5.0553	&0.2133 \\
89	&35	&1.1089	&5.1943	&0.2135	&151	&59	&1.1352	&5.1943	&0.2185 \\
88	&35	&1.0992	&5.3738	&0.2045	&152	&59	&1.2020	&5.3738	&0.2237 \\
87	&34	&1.1536	&5.2821	&0.2184	&153	&60	&1.1393	&5.2821	&0.2157 \\
86	&34	&1.1251	&5.3757	&0.2093	&154	&60	&1.1868	&5.3757	&0.2208 \\
85	&34	&1.1045	&5.5223	&0.2000	&155	&60	&1.2470	&5.5223	&0.2258 \\
84	&33	&1.1650	&5.4366	&0.2143	&156	&61	&1.1849	&5.4366	&0.2179 \\
83	&33	&1.1594	&5.6605	&0.2048	&157	&61	&1.2619	&5.6605	&0.2229 \\
82	&33	&1.1391	&5.8380	&0.1951	&158	&61	&1.3302	&5.8380	&0.2278 \\
81	&32	&1.1577	&5.5159	&0.2099	&159	&62	&1.2142	&5.5159	&0.2201 \\
80	&33	&1.0730	&6.1315	&0.1750	&160	&61	&1.4562	&6.1315	&0.2375 \\
\end{tabular}
\end{ruledtabular}
\end{table*}

\subsection{Deformation Parameter}
It is widely acknowledged that  the shapes of the primary fragments at the scission moment are hardly obtained not only by the measurements but also by the theories. In order to obtain the general descriptions of the deformation parameters of the primary fragment pairs, average total kinetic energies are assumed to be only provided by the Most Probable Primary Fragment Pairs, and the effects of the excited energies on the deformation parameters are ignored at the scission moment because of the large deformation energies. Therefore, it is presumed that the deformation parameters of the primary fragments are dependent on the isospin asymmetry degree $I$, and can be expressed as:
\begin{equation}
\label{beta}
\beta_i=\alpha_i I_i,
\end{equation}
where $\alpha_i$ is the phenomenological parameter, and $I_i=(N_i-Z_i)/A_i$ ($i=L0$ or $H0$).

\begin{figure}
\includegraphics[width=8.5cm]{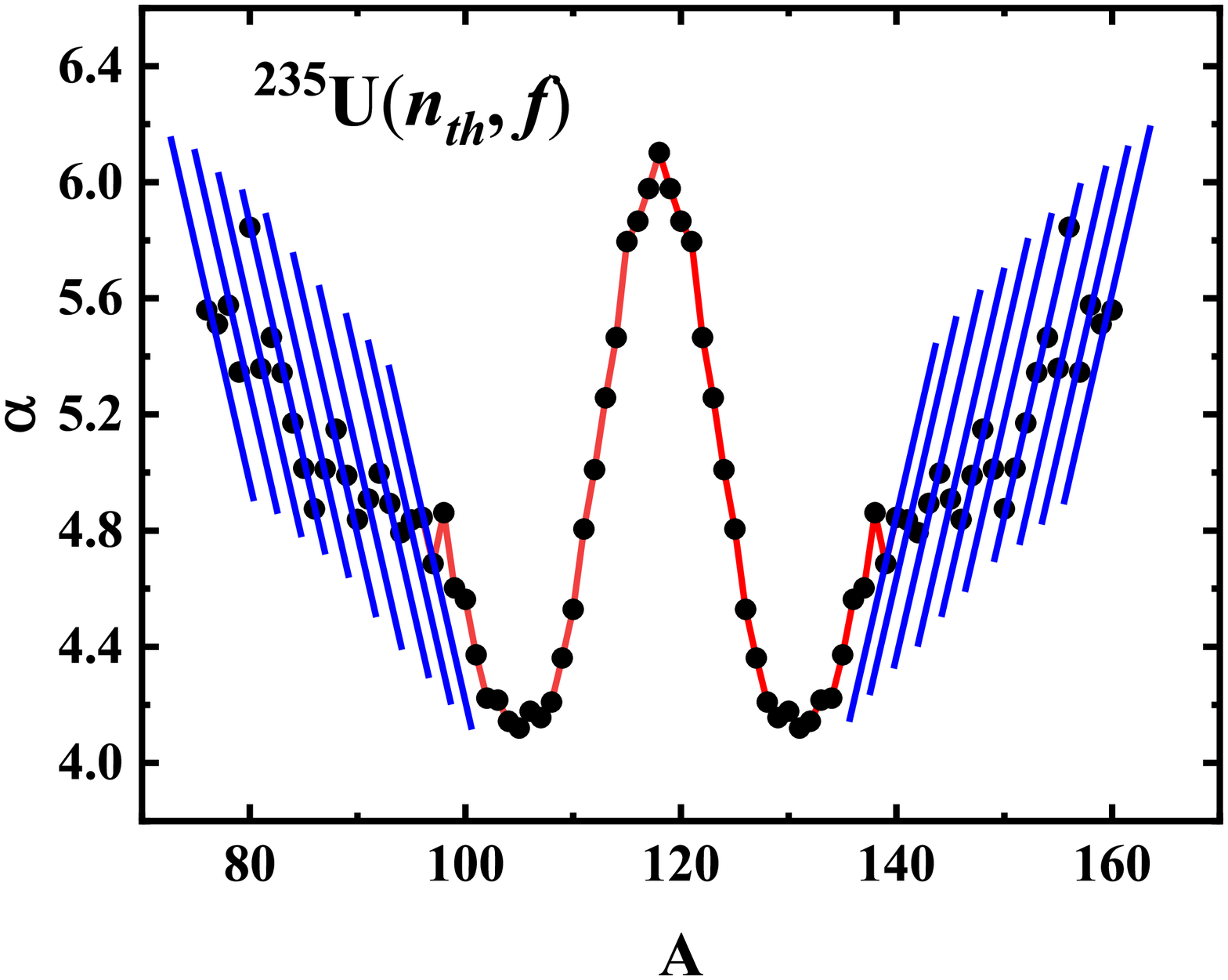}
\caption{\label{fig:alpha0}(Color online) The relationship between phenomenological parameters $\alpha$ and the mass number $A$ for $^{235}$U$(n_{th}, f)$ reaction. The red and blue lines show the results of Eqs. (\ref{alpha01})- (\ref{alpha03}).}
\end{figure}

\begin{figure}
\includegraphics[width=8.5cm]{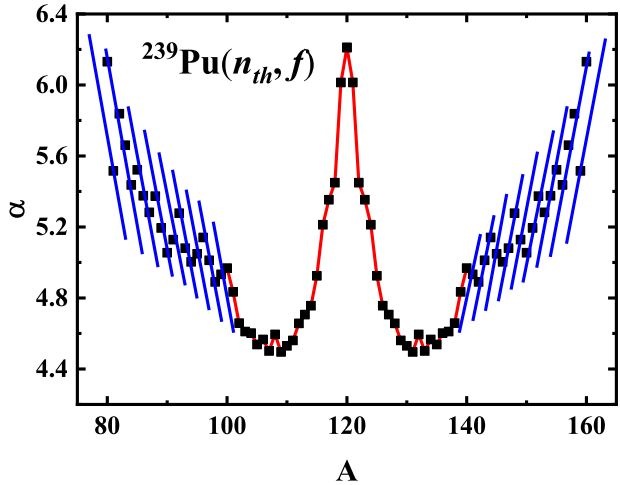}
\caption{\label{fig:alpha0_Pu}(Color online) The relationship between phenomenological parameters $\alpha$ and the mass number $A$ for $^{239}$U$(n_{th}, f)$ reaction.}
\end{figure}

By substituting Eq.~(\ref{beta}) into Eq.~(\ref{atke}), the relationship between the mass number $A$ and the phenomenological parameter $\alpha_i$ can be obtained, as shown in Figs. \ref{fig:alpha0} and \ref{fig:alpha0_Pu}, from which the same slopes of the isotopes are seen as the blue lines in the range $A\ge 140$, and $A\le A_f-140$ with negative slopes. And in the residual region, the phenomenological parameter $\alpha_i$ is approximately viewed as the smooth single-valued function of the mass number $A$. The critical point $A=140$ is exactly corresponded to the peaks of the measured mass distributions of the heavy fragments for low energy neutron-induced fission of actinides \cite{Baba1997, Tsuchiya2000, Thierens1984, Sun2012, Sun2014, Sun2015}. 

A large amount of the measured average total kinetic energy distributions, such as $^{231}$Pa, $^{232}$U, $^{233}$U, $^{235}$U, $^{237}$Np, $^{239}$Pu, $^{241}$Pu, $^{241}$Am and $^{243}$Am induced by thermal neutrons, show that they peak at $A \approx 132$ \cite{Asghar1978, Asghar1981, Baba1997, Wagemans1981, Tsuchiya2000, Thierens1984, Asghar1980}. And many measured charge distributions show that they peak at $Z\approx 54$ \cite{Reisdorf1971, Naik1997, Naik2004}. It is widely acknowledged that $A \approx 132$ and $Z\approx 54$ are closely related to the shell structure. It is inspired that  $A \approx 132, 140, A_f/2$ and $Z\approx 54, Z_f/2$ are some crucial values for the average total kinetic energy distributions for thermal neutron-induced fission of actinides. Therefore, the expression of the phenomenological parameter $\alpha$ can be presumed as:
\begin{subequations}
\begin{equation}
\label{alpha01}
\begin{aligned}
\alpha=a_{L,~H}+a_k(A-A_i^p)^2,\\
(Z_f-54)\le Z \le Z_f/2~or~Z_f/2 \le Z \le 54,
\end{aligned}
\end{equation}
and
\begin{equation}
\label{alpha02}
\begin{aligned}
\alpha=\delta_{L,~H}bZ+c_{L,~H}+\delta_{L,~H}kA,\\
Z\le (Z_f-54)~or~Z\ge 54.
\end{aligned}
\end{equation}
\end{subequations}
Where $A_{H0}^p$=132, $A_{L0}^p=A_f-A_{H0}^p$ and $\delta$$_L$=-1, $\delta_H$=1. If the deformation parameter $\beta$ is assumed as a smooth function of $A$ and $Z$ for a large amount of the primary fragments, so the coefficients in Eqs.~(\ref{alpha01}) and (\ref{alpha02}) can be derived as:
\begin{equation}
\label{alpha03}
\left\{
\begin{array}{l}
a_{k}=k/16 \\
a_{L}=-(Z_{f}-54)b+c_L-(A_{f}-136)k \\
a_{H}=54 b+c_{H}+136k \\
c_{H}=-Z_{f}b-A_{f} k+c_L.
\end{array}\right.
\end{equation}
Thus, there are only three adjustable free coefficients \{$k,~b,~c_L$\}  for describing the general rule of the deformation parameter $\beta$ of the Most Probable Primary Fragment Pairs. Furthermore, this rule is extended to describe the deformation parameters of arbitrary primary fragments for $^{235}$U$(n_{th}, f)$ and $^{239}$Pu$(n_{th}, f)$ reactions. 

\begin{figure}
\includegraphics[width=8.5cm]{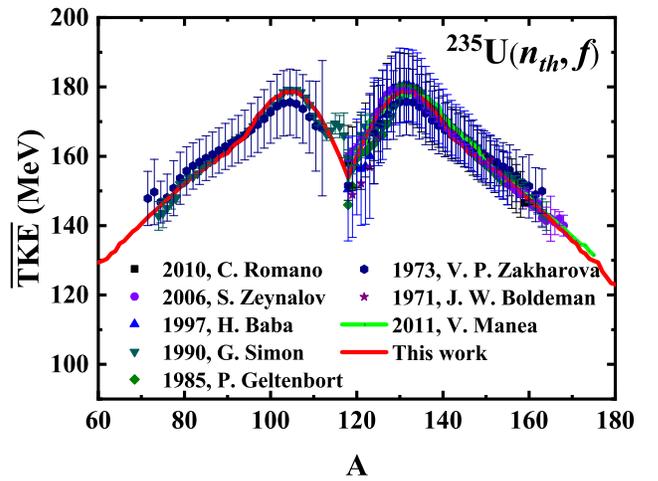}
\caption{\label{fig:tke235}(Color online) Comparisons of the calculated average total kinetic energy $\overline{TKE}(A)$ with the measurements for $^{235}$U($n_{th}, f$) reaction. The experimental data are derived from Refs. \cite{ Baba1997,  Simon1990, Zeynalov2006, Geltenbort1986, Zakharova1973, Boldeman1971}. The red line and green lines denote the theoretical results of this work and PSC approach \cite{Manea2011}, respectively.}
\end{figure}

\begin{figure}
\includegraphics[width=8.5cm]{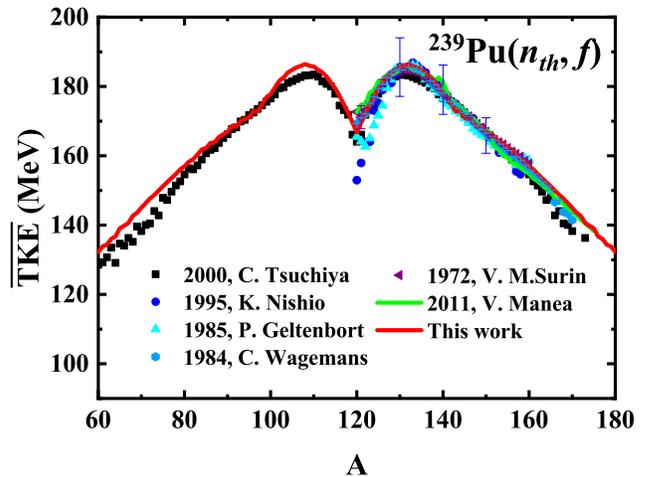}
\caption{\label{fig:tke239}(Color online)  Comparisons of the calculated average total kinetic energy $\overline{TKE}(A)$ with the measurements for $^{239}$Pu($n_{th}, f$) reaction. The experimental data are derived from Refs. \cite{Tsuchiya2000, Geltenbort1986, Wagemans1984, Nishio1995, Surin1971}.}
\end{figure}

\begin{figure}
\includegraphics[width=8.5cm]{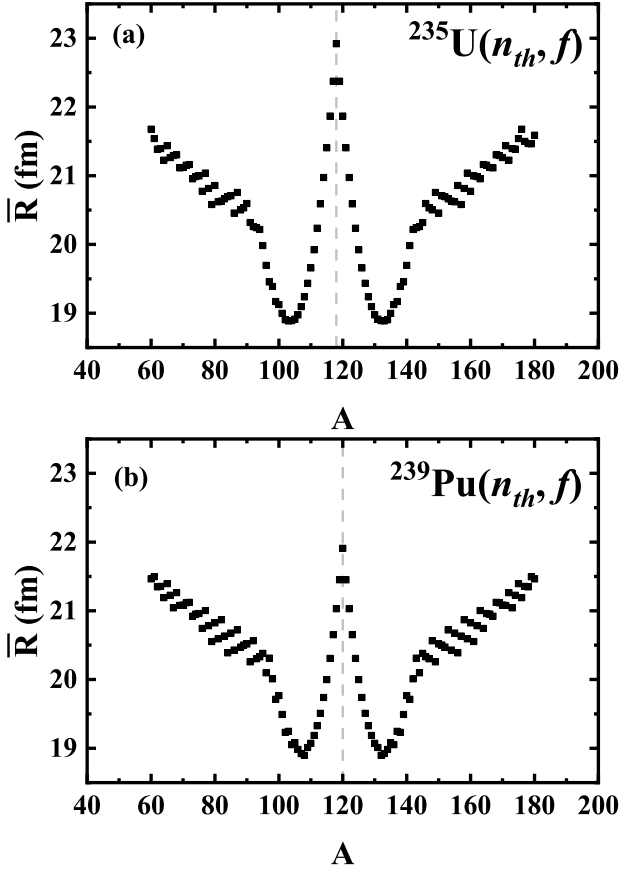}
\caption{\label{fig:car}(Color online) The average distance $\overline{R}$ versus mass number $A$ for $^{235}$U$(n_{th},f)$ reaction (a) and  $^{239}$Pu$(n_{th},f)$ reaction (b).}
\end{figure}

\begin{figure}
\includegraphics[width=8.5cm]{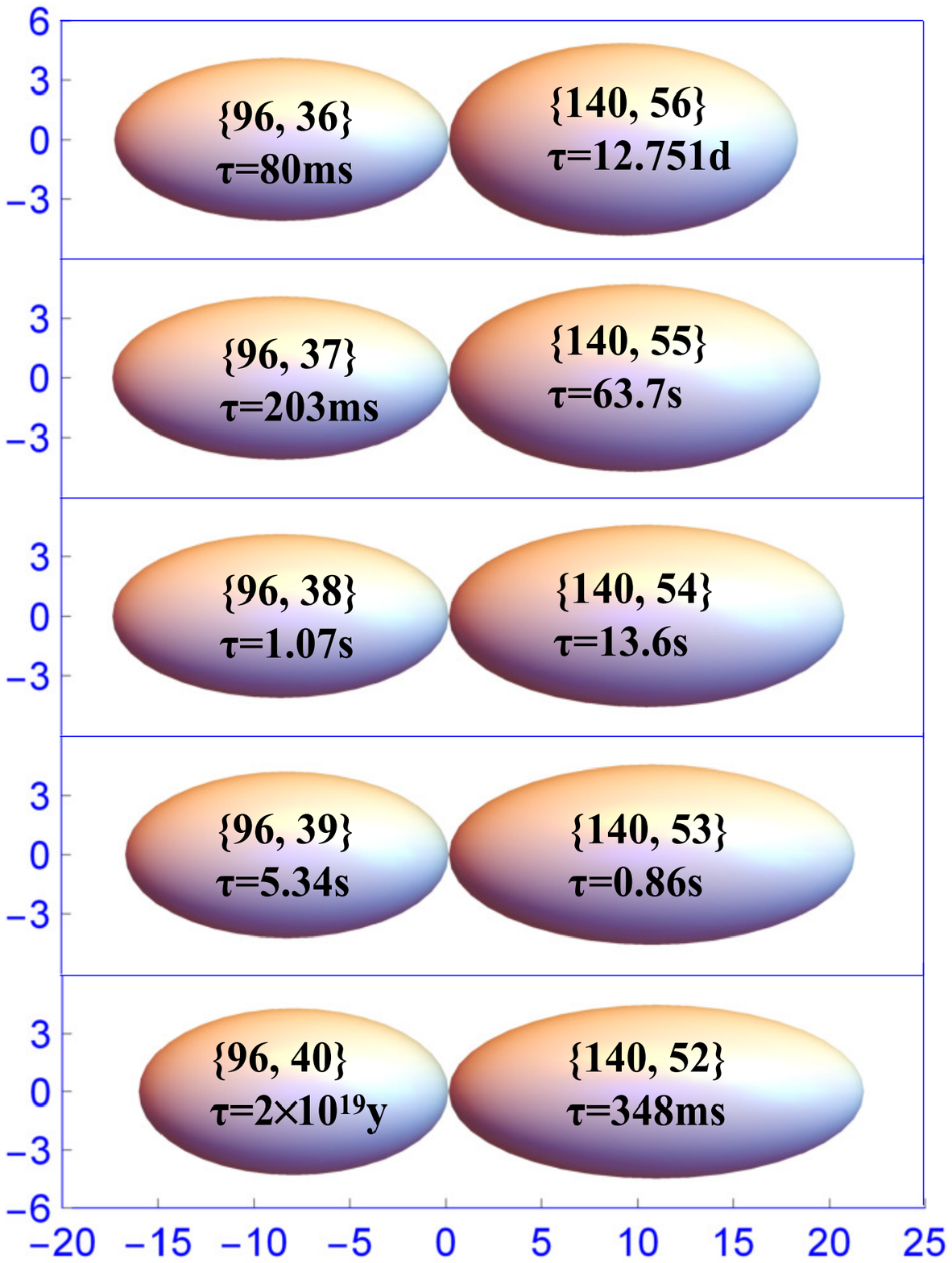}
\caption{\label{fig:tou}(Color online) The deformation at the scission moment and half-life in the ground state of the possible isobar pairs of $A_{L0}=96$ and $A_{H0}=140$ for $^{235}$U($n_{th},~f$) reaction.}
\end{figure}

\begin{figure}
\includegraphics[width=8.5cm]{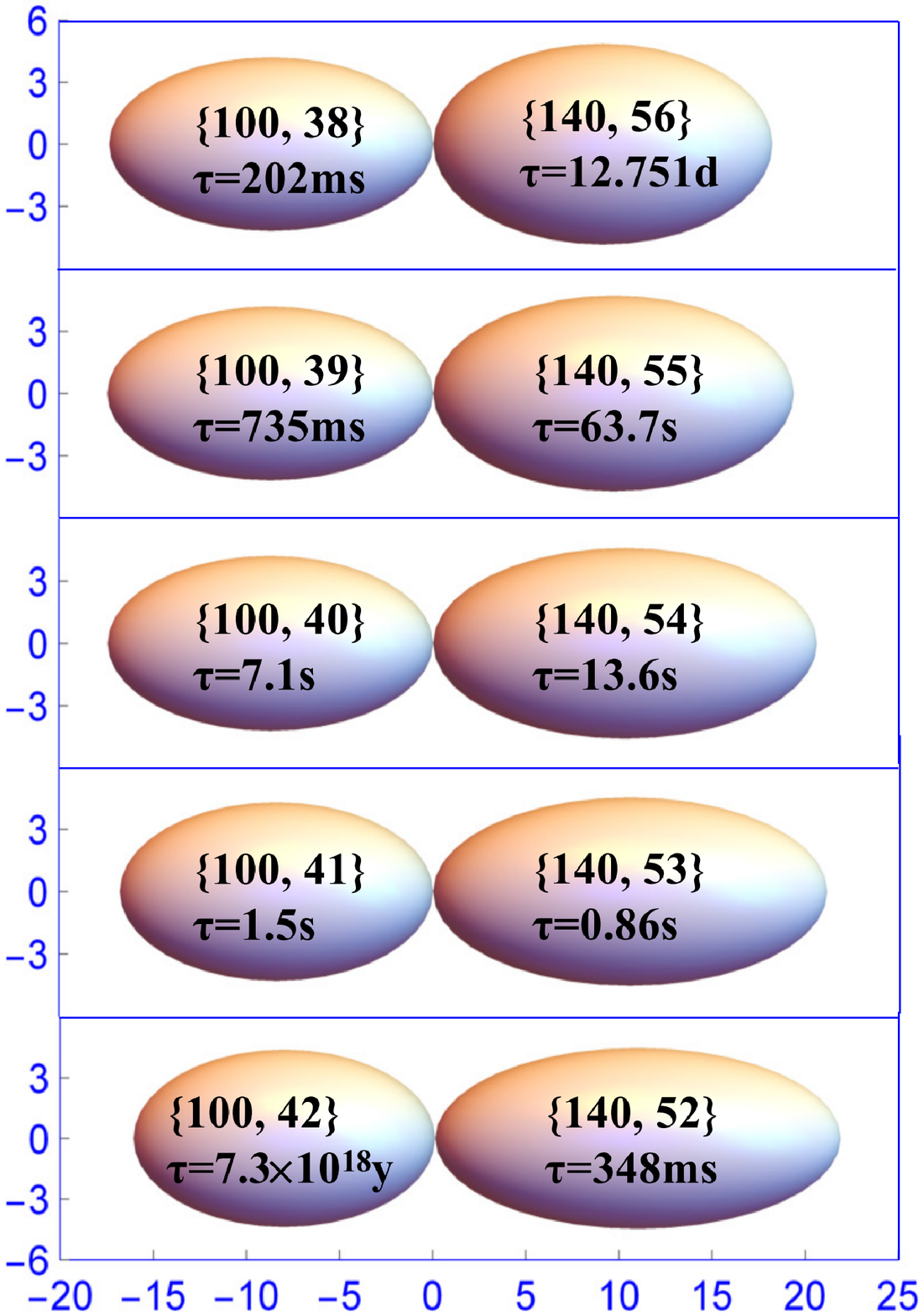}
\caption{\label{fig:tou_Pu}(Color online)  The deformation at the scission moment and half-life in the ground state of the possible isobar pairs of $A_{L0}=100$ and $A_{H0}=140$ for $^{239}$Pu($n_{th}, f$) reaction.}
\end{figure}

\begin{figure}
\includegraphics[width=8.5cm]{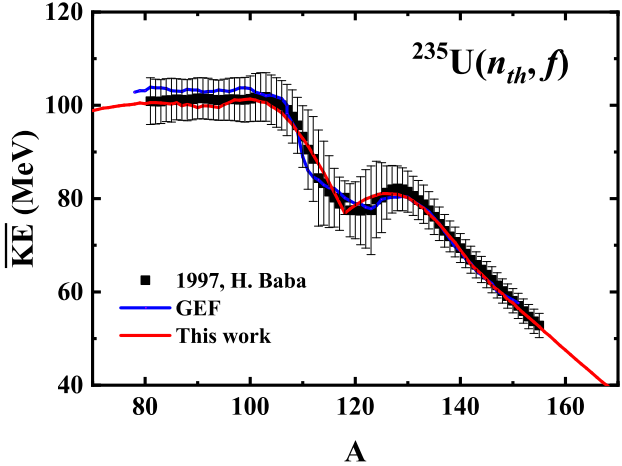}
\caption{\label{fig:ke235}(Color online) Comparisons of the average kinetic energy $\overline{KE}(A)$ with the measurements and model calculations for $^{235}$U($n_{th}, f$) reaction. The experimental data are derived from Ref. \cite{Baba1997}. The red and blue lines denote the results of this work and GEF model \cite{Schmidt2016}, respectively.}
\end{figure}

\begin{figure}
\includegraphics[width=8.5cm]{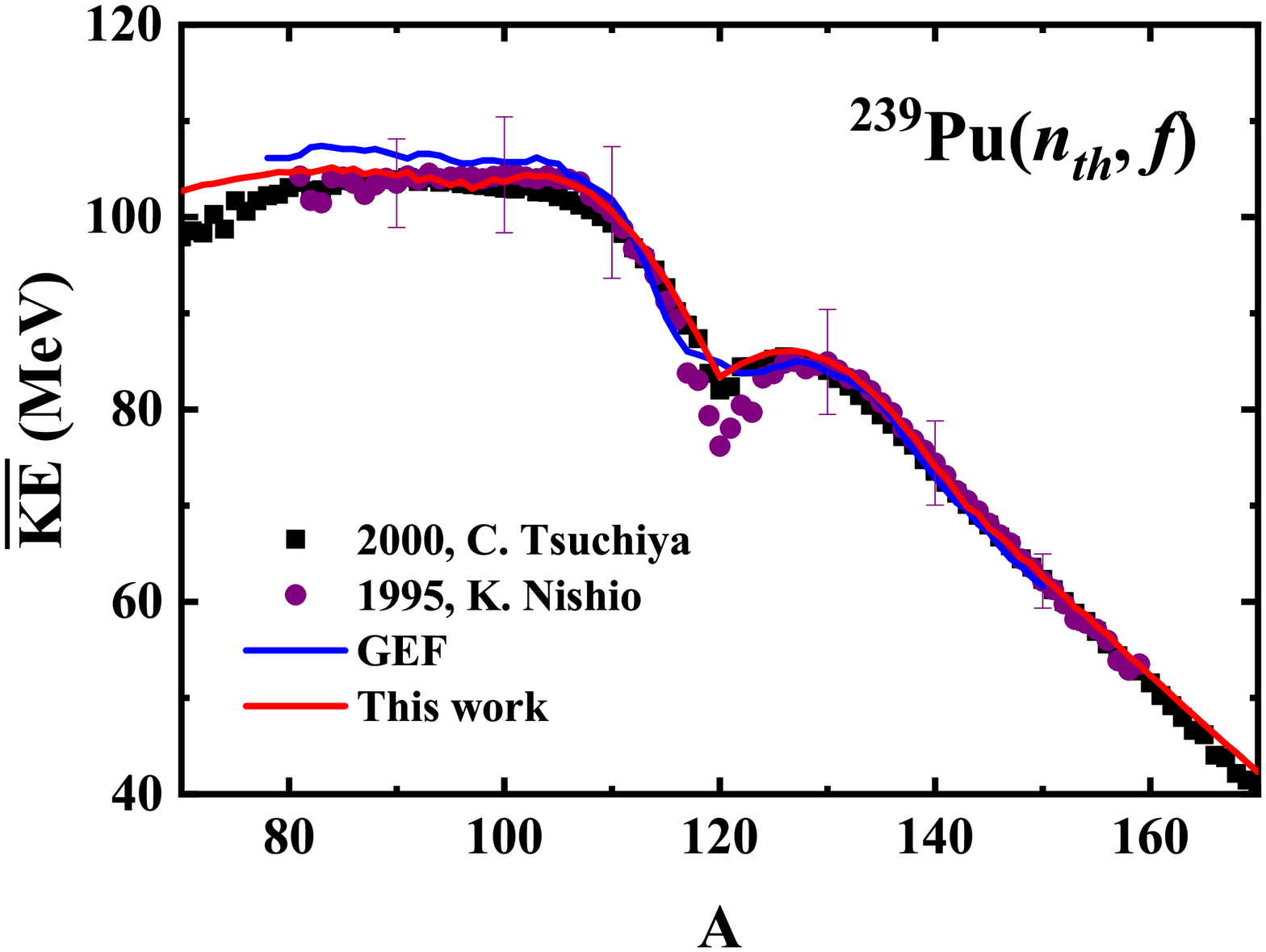}
\caption{\label{fig:ke239}(Color online) Comparisons of the average kinetic energy $\overline{KE}(A)$ with the measurements and model calculations  for $^{239}$Pu($n_{th}, f$) reaction. The experimental data are derived from Refs. \cite{Tsuchiya2000, Nishio1995}.}
\end{figure}

\section{Result and analysis}\label{sec4}
\subsection{Average total kinetic energy}
Based on the DSM introduced in the last sections II and III, the explicit average total kinetic energy distribution $\overline{TKE}(A)$ is expressed as Eq. (\ref{atke}), and the quadrupole deformation parameter of the arbitrary primary fragments is also expressed as Eqs. (\ref{beta})-(\ref{alpha03}). Thus, the root-mean-square deviation $\sigma$ can be expressed as:
\begin{equation}
\sigma=\sqrt{\frac{1}{\sum_j}\sum_j[\overline{TKE}_{th}(A_j)-\overline{TKE}_{exp}(A_j)]^2},
\end{equation}
which is adopted to determine the agreement between the theoretical calculations and the experimental data. Where the $\overline{TKE}_{th}$ and $\overline{TKE}_{exp}$ donote the theoretical values and experimental data with small relative errors \cite{Baba1997, Tsuchiya2000} are selected to obtain a set of the optimal parameters $\{k,~b,~c_L \}$ listed in Table \ref{tab:cs} both for $^{235}$U$(n_{th},f)$ and  $^{239}$Pu$(n_{th},f)$ reactions. 

Figs. \ref{fig:tke235} and \ref{fig:tke239} show the comparisons of the calculated average total kinetic energies $\overline{TKE}(A)$ with the measurements for $^{235}$U($n_{th}, f$) and $^{239}$Pu($n_{th}, f$) reactions, respectively. The experimental data are derived from Refs. \cite{Simon1990, Baba1997, Zeynalov2006,  Geltenbort1986, Zakharova1973, Boldeman1971, Wagemans1984, Nishio1995, Surin1971}. The red line and green lines denote the calculated results of this work and Pre-Scission Configuration (PSC) approach \cite{Manea2011}, respectively. It can be seen that all of the calculated results are reasonable, with their root-mean-square deviations $\sigma_{\overline{TKE}}$ listed in Table \ref{tab:sigma}. It is obvious that the results of this work is slightly superior to those of the PSC approach.

\begin{table}
\caption{\label{tab:cs}The optimal coefficients of the deformation parameter (no units).}
\begin{ruledtabular}
\begin{tabular}{ccc}
$k$&$b$&$c_L$ \\\hline
0.1943&-0.4151&7.7924\\
\end{tabular}
\end{ruledtabular}
\end{table}

\begin{table}
\caption{\label{tab:sigma}Root-mean-square deviations $\sigma$ of the different models.}
\begin{ruledtabular}
\begin{tabular}{rccl}
Fission system&$\sigma_{\overline{TKE}}$ (MeV)&$\sigma_{\overline{KE}}$ (MeV)&Models \\
\hline
$^{235}$U($n_{th},~f$) &1.32&1.26&DSM         \\
                       &2.01&/   &PSC \cite{Manea2011}\\
                       &/   &1.73&GEF \cite{Schmidt2016}\\
$^{239}$Pu($n_{th},~f$)&2.11&1.09&DSM\\
                       &2.95&/   &PSC \cite{Manea2011}\\
                       &/   &2.22&GEF \cite{Schmidt2016}\\
\end{tabular}
\end{ruledtabular}
\end{table}

From Eq. (\ref{Vc1}), it can be seen that the distance $R$ between the mass centers of the complementary light and heavy primary fragments is critical. In terms of Eq. (\ref{R}), the average distance $\overline{R}$ between the mass centers of complementary primary fragments can also be rewritten as a function of mass number $A$: 
\begin{equation}
\begin{aligned}
\overline{R}(A)=&\frac{1}{\sum_k}\sum_{k}R(A_{L0}(j), Z_{L0}(j, k),\\
&\beta_{L0}(j, k); A_{H0}(j), Z_{H0}(j, k), \beta_{H0}(j, k)).
\end{aligned}
\end{equation}

Fig. \ref{fig:car} shows the average distance $\overline{R}$ versus the mass number $A$ for $^{235}$U$(n_{th},f)$ reaction (a) and  $^{239}$Pu$(n_{th},f)$ reaction (b), from which it can be seen that there are one peak at the symmetric fission point $A_f/2$ and two valleys at $A=132$ and $A=A_f-132$ positions, with the peak value for $^{235}$U$(n_{th},f)$ reaction being higher than that for $^{239}$Pu$(n_{th},f)$ reaction. It implies that the $\overline{TKE}$ at the symmetric fission point $A_f/2$ of $^{235}$U$(n_{th},f)$ reaction is smaller than that of $^{239}$Pu$(n_{th},f)$ reaction, as shown in Figs. \ref{fig:tke235} and \ref{fig:tke239}.

Fig. \ref{fig:tou} shows the deformation at the scission moment and half-life in the ground state of the possible isobar pairs of $A_{L0}=96$ and $A_{H0}=140$ for $^{235}$U($n_{th},~f$) reaction. And Fig. \ref{fig:tou_Pu} shows the deformation at the scission moment and half-life in the ground state of the possible isobar pairs of $A_{L0}=100$ and $A_{H0}=140$ for $^{239}$Pu($n_{th},~f$) reaction. Obviously, the isobar pair with the smallest half-life in the ground state holds the smallest deformation at the scission moment. This implies that the kinetic energy of the primary fragments can be manifested by some properties of their ground state. Several properties of the Most Probable Primary Fragment Pairs, such as the quadrupole deformation parameter $\beta_i$, phenomenological parameter $\alpha_i$ and the isospin asymmetry degree $I_i$, are listed in Tables \ref{tab:beta} and \ref{tab:beta_Pu} for $^{235}$U($n_{th},~f$) and  $^{239}$Pu($n_{th},~f$) reactions, respectively.

\subsection{Average kinetic energy}
The subjects of the total kinetic energy $TKE$ are the complementary primary fragment pairs. However, the energy allocation of light and heavy primary fragments is different. In terms of the conservations of momentum and kinetic energy as follows, 
\begin{equation}
\left\{
\begin{array}{l}
m_{L0}v_{L0}=m_{H0}v_{H0} \\
\frac{1}{2}m_{L0}v^2_{L0}+\frac{1}{2}m_{H0}v^2_{H0}=TKE,
\end{array}\right.
\end{equation}
the kinetic energy of the heavy primay fragment $KE_{H0}$ is defined as:
\begin{equation}
\begin{aligned}
KE_{H0}&=\frac{1}{2}m_{H0}v^2_{H0}\\
&=\frac{m_{L0}}{m_{L0}+m_{H0}}TKE\\
& \approx \frac{A_{L0}}{A_f}TKE.
\end{aligned}
\end{equation}
And the kinetic energy of the light primary fragment $KE_{L0}$ can be easily written as:
\begin{equation}
\begin{aligned}
KE_{L0}&=TKE-KE_{H0}\\
&\approx \frac{A_{H0}}{A_f}TKE.
\end{aligned}
\end{equation}
Thus, the average kinetic energy distribution $\overline{KE}(A)$ can be expressed as:
\begin{equation}
\label{eq_KE1}
\begin{aligned}
\overline{KE}(A_{})=&\frac{1}{{\sum_k}}\sum_{k}KE(A_{L0}(j), Z_{L0}(j, k);\\ & A_{H0}(j), Z_{H0}(j, k)),
\end{aligned}
\end{equation}
or
\begin{equation}
\label{eq_KE2}
\overline{KE}(A)=\frac{A_f - A}{A_f} \overline{TKE}(A).
\end{equation}
Further, it can be found that the results of Eqs. (\ref{eq_KE1}) and (\ref{eq_KE2}) are roughly equal. Evidently, Eq. (\ref{eq_KE2})
 simple and convenient, is used in this paper.

Figs. \ref{fig:ke235} and \ref{fig:ke239} show the comparisons of the calculated and measured $\overline{KE}(A)$ for $^{235}$U($n_{th}$, $f$) and $^{239}$Pu($n_{th}$, $f$), respectively.
The red and blue lines denote the results of this works and GEF model \cite{Schmidt2016}, respectively. And the root-mean-square deviations $\sigma_{\overline{KE}}$ of this work and GEF are listed in Table \ref{tab:sigma}. From Figs. \ref{fig:ke235} and \ref{fig:ke239} and Table \ref{tab:sigma}, it can be seen that all of the calculated results of this paper is reasonable and slightly superior to those of GEF model.

\section{Conclusion}\label{sec5}

Based on the dinuclear concept, if it is assumed that the fissile nucleus elongates along an coaxis and attains the scission point, the kinetic energy of the complementary primary fragment pairs is mainly provided by the Coulomb repulsion. After performing strictly the six-dimensional integral and omitting the higher order terms, the compact expression of the Coulomb repulsion is explicitly obtained. In terms of the statistical properties of the abundant experimental total kinetic energy distributions, mass and charge distributions, and our previous theoretical results, some special quantities (such as mass numbers $A=132$ and 140, charge number $Z=54$, and symmetrical fission point \{$A_f/2, Z_f/2$\}) are used to derive the phenomenological expression of the deformation parameters of the different primary fragments at the scission moment. And a set of optimal coefficients $\{k, b, c_L\}$ of the deformation parameters of arbitrary complementary fragments is obtained both for $^{235}$Pu($n_{th}, f$) and $^{239}$Pu($n_{th}, f$) reactions. The calculated results concurrently  agree well with the measured total kinetic energy distributions $\overline{TKE}(A)$ and experimental kinetic energy distributions $\overline{KE}(A)$, and are also slightly better than the previous theoretical ones.

However, the DSM model omits the rotational energy, which largely contributes to heavy-ion induced reactions due to the big angular momenta of the incident particle. In addition, nuclear charge densities have an exponential tails rather than a sharp surface, so DSM ignores the attractive nuclear force at the scission moment because of the super coaxial deformation. Due to the fact that the DSM model can provide enough trust for the $\overline{TKE}(A)$ and $\overline{KE}(A)$, the feasibility is expected to extend the arbitrary incident energies and/or other fissile systems.

\begin{center}
{\bf  ACKNOWLEDGMENTS}
\end{center}
We appreciate the valuable suggestions of our colleagues Drs. Ning Wang, Li Ou and Min Liu.

\nocite{*}

\bibliography{apssamp}

\end{document}